\documentclass[reprint,amsfonts,amsmath,amssymb,aps,]{revtex4-1}

\usepackage{graphicx}
\usepackage{dcolumn}
\usepackage{bm}

\def\be{\begin{equation}}
\def\ee{\end{equation}}

\def\be{\begin{equation}}
\def\ee{\end{equation}}

\def\bg{\bar{g}}
\def\beq{\begin{eqnarray}}\def\eeq{\end{eqnarray}}
\def\ba#1\ea{\begin{align}#1\end{align}}
\def\bg#1\eg{\begin{gather}#1\end{gather}}
\def\bm#1\em{\begin{multline}#1\end{multline}}
\def\bmd#1\emd{\begin{multlined}#1\end{multlined}}

\def\({\left(}
\def\){\right)}
\def\[{\left[}
\def\]{\right]}

\newcommand{\pd}{\partial}
\newcommand{\bea}{\begin{eqnarray}}
\newcommand{\eea}{\end{eqnarray}}
\newcommand{\een}{\end{enumerate}}
\newcommand{\bi}{\begin{itemize}}
\newcommand{\ei}{\end{itemize}}
\newcommand{\bc}{\begin{center}}
\newcommand{\ec}{\end{center}}
\newcommand{\bfig}{\begin{figure}}
\newcommand{\efig}{\end{figure}}
\newcommand{\bt}{\begin{table}}
\newcommand{\et}{\end{table}}
\newcommand{\btab}{\begin{tabular}}
\newcommand{\etab}{\end{tabular}}
\newcommand{\bs}{\begin{slide}}
\newcommand{\es}{\end{slide}}

\begin{document}

\preprint{WITS-CTP-155}

\title{Renormalized Entanglement Entropy for BPS Black Branes}

\author{Arpan Bhattacharyya}
\email{arpan@cts.iisc.ernet.in}
\affiliation{Centre for High Energy Physics, Indian Institute of Science, Bangalore 560012, India}
\author{S.\ Shajidul Haque}
\email{Shajid.Haque@wits.ac.za}
\author{\'{A}lvaro V\'{e}liz-Osorio}
\email{Alvaro.VelizOsorio@wits.ac.za}
\affiliation{Centre for Theoretical Physics, NITheP, and School of Physics, University of the Witwatersrand, Johannesburg, WITS 2050, South Africa}

\date{\today}

\begin{abstract}
We compute the renormalized entanglement entropy (REE) for BPS black solutions in ${\cal N}=2$, 4d gauged supergravity.
We find that this quantity decreases monotonically with the size of the entangling region until it reaches a critical point, then increases and  approaches the entropy density of the brane. This behavior can be understood as a consequence of the REE being driven by two competing factors, namely entanglement and the mixedness of the black brane. In the UV entanglement dominates, whereas in the IR the mixedness takes over. 

\end{abstract}

\maketitle

\section {Introduction}
Over the last decade there has been great interest in the study of entanglement entropy (EE) for a wide variety of quantum systems. This quantity measures the amount of entanglement between two subsystems. One of the key advances that motivated these developments was the observation by Ryu and Takayanagi \cite{Ryu:2006bv} that EE can be computed holographically. This quantity is sensitive to UV physics and the leading divergence obeys the area law. This indicates that the EE is not a well defined observable in the continuum limit. 

In 2d, Casini and Huerta \cite {Casini:2012ei} devised a method to extract the universal contribution to the entanglement entropy. Liu and Mezei \cite {Liu1,Liu2} generalized this prescription to higher dimensions. The resulting quantity, known as the renormalized entanglement entropy (REE), 
is UV finite and local on the scale of the entangling region. The evolution of REE with respect to the size of the entangling region can be used as a probe to realize the RG flow. Moreover, for vacuum states the REE for a spherical entangling region provides a c-function parametrizing the RG flow. We want to compute the REE for states that break Lorentz invariance due to the presence of a finite charge density.  While holographic RG flows for vacuum states correspond to domain walls (i.e. solutions interpolating between same dimensional $AdS$) \cite {Girardello:1999bd, Freedman:1999gp, Myers:2010tj}, the flow for charged states can be described in terms of black holes/branes. 

Specifically, we want to study the REE for BPS black solutions in ${\cal N}=2$, 4d FI gauged supergravity. Starting with the work of Cacciatori-Klemm \cite {Cacciatori:2009iz} these models have been studied extensively over the last few years \cite {Dall'Agata:2010gj, Barisch:2011ui, Barisch-Dick:2013xga}, culminating with the full construction of static BPS solutions for all symmetric models by \cite {Katmadas, Halmagyi:2014qza}. These are solutions that  interpolate between $AdS_4$ and $AdS_2 \times \Sigma_k$, where $\Sigma_k$ is the surface of constant sectional curvature with $k= -1,0,1$. Since these objects interpolate between different AdS spaces they are interesting from the holographic perspective. Morever, for the STU model these solutions have a  M-theory realization through an embedding into the de Wit-Nicolai ${\cal N}=8$ theory \cite{de Wit:1986iy}. 

This paper is organized as follows. In section \ref{REEa} we outline the computation for the REE for the black brane solutions. In section \ref{BPS}, we summarize the BPS black objects in $AdS_4$, followed by the computation for the REE in section \ref{branes}. Finally, in  Appendix \ref {ApA}	, we discuss the symplectic invariant in ${\cal N}=2$.

%%%%%%%%%%%%%
%%%%%%%%%%%%%
%%%%%%%%%%%%%

\section{Renormalized Entanglement Entropy}\label{REEa}
\begin{figure}[h!]
  \centering
\includegraphics[trim=1cm 1cm 1cm 1cm,scale=0.55]{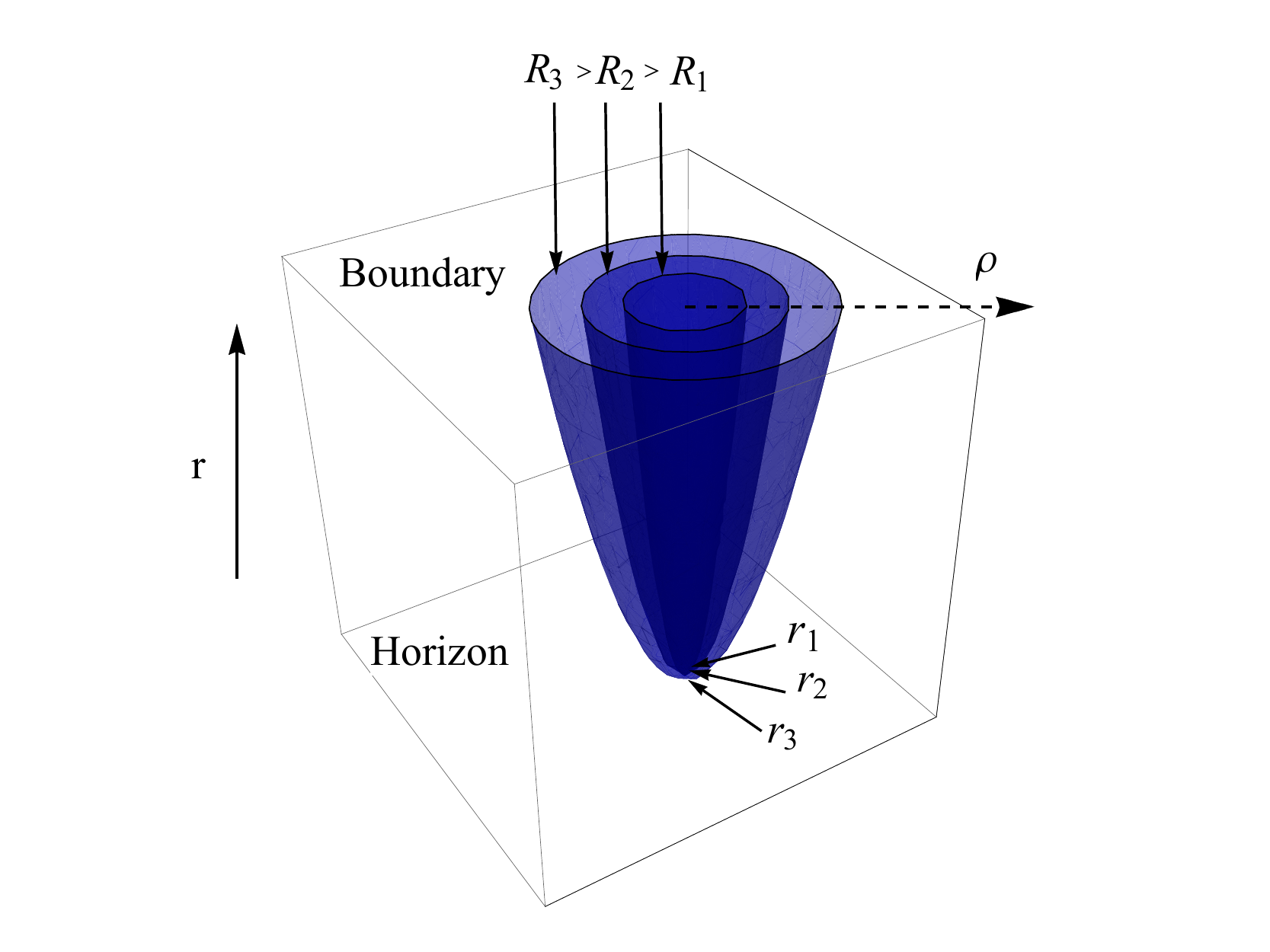}
  \caption{Minimal surfaces in $AAdS$}\label{minimal figure}
\end{figure}

In this section we outline a general procedure to obtain the universal contributions to the entanglement entropy for 
quantum systems that can be described holographically by a metric of the form
\begin{equation}\label{metric ansatz}
	ds^2= -a^2 dt^2+a^{-2} dr^2+b^2 d\Omega_k^2. 
\end{equation}
In the above expression, $d\Omega_k^2$ is the line element of a surface of constant sectional curvature $k$. Clearly we must demand AdS asymptotics. Hence, as $r\rightarrow\infty$ the metric takes the form
\begin{equation}
a\rightarrow\frac{r}{l_4}\hspace{9mm}b\rightarrow\frac{r}{l_4}.
\end{equation}
We wish to compute the entanglement entropy for a subsystem $A$ consisting of a disk $\Sigma(R)$ of radius $R$. From the Ryu-Takayanagi prescription \cite{Ryu:2006bv}, we know that this quantity corresponds to the area of an extremal surface attached to $\pd \Sigma(R)$  going into the bulk (see Fig. \ref{minimal figure}). For the metric (\ref{metric ansatz}) this problem corresponds to the Plateau problem for the functional 
\begin{equation}\label{area functional}
S(R)=\frac{2 \pi}{l_p^2}\int_0^Rd\rho\,\rho\,b^2\sqrt{1+e^{-2\psi}\dot r^2}\, \, \hspace{9mm} e^\psi\equiv ab.
\end{equation}
In the subsequent calculations we will absorb the factor $\frac{2 \pi } {l_p^2}$ into $S(R)$. From the above functional, it can be showed that the profile $r(\rho)$ of the minimal surfaces can be found from the ODE
\begin{equation}\label{extremal}
\ddot r-\psi' \;  \dot r^2+\left(\frac{\dot r}{ \rho}-2\frac{b'}{b} e^{2\psi}\right)\left(1+e^{-2\psi}\,\dot r^2\right)=0,
\end{equation}
where $\dot\,=\pd_\rho$ and $ '=\pd_r$.
This equation is supplemented with the boundary conditions 
\begin{equation}\label{boundary condition}
r(0)= r_{0}>0\hspace{9mm} \dot r(0)=0.
\end{equation}
Solutions of equations (\ref{extremal}) and (\ref{boundary condition}) correspond to extremal surfaces attached to the boundary of a disk at infinity and whose tip is at $r=r_0$. Moreover, the depth of the tip can be related to the size $R$ of the entangling disk at the boundary, $R$ corresponds to the value of $\rho$ for which $r(\rho)\rightarrow\infty$. Therefore, each surface can be labeled either by the size of the entangling region $R$ or by the depth it reaches in the bulk $r_0$  (see Fig. \ref{minimal figure}).

Once we have found the profile $\tilde r(\rho)$ of the minimal surface, we are instructed to plug it into the functional (\ref{area functional}) in order to obtain the holographic entanglement entropy. However, one must be careful since the resulting quantity is divergent. 
We should, therefore, regularize this integral first. We introduce a UV cut-off in the following way--let $\epsilon\ll 1$  and 
restrict the values of $r$ such that 
\be \label{reg}
\frac{l_4}{b(r)}>\epsilon .
\ee
 Using $\tilde r(\rho)$, we can translate this bulk cut-off into a boundary cut-off, i.e., we must consider only $\rho<R_\epsilon$, where
\be \label{reg2}
\frac{l_4}{b(\tilde r(R_\epsilon))}= \epsilon .
\ee
Then we can compute the finite quantity
\begin{equation}\label{area functional reg}
S(R,\epsilon)=\int_0^{R_{\epsilon}} d\rho\,{\cal A}(\rho,R),
\end{equation}
where ${\cal A}(\rho,R)$ stands for the integrand of (\ref{area functional}) evaluated on the solution $\tilde r(\rho)$.

In order to systematically obtain the universal contribution to (\ref{area functional reg}), which we call hereafter renormalized entanglement entropy, we use the operator introduced in  \cite {Casini:2012ei, Liu1}. In four bulk space-time dimensions, the renormalized entanglement entropy is given by 
\begin{equation}
{\cal S}(R)\equiv \left(R\frac{d}{dR}-1\right)S(R,\epsilon).
\end{equation}
This quantity can be alternatively written as 
\begin{equation}\label{ree}
{\cal S}(R)={\cal A}(R,R_\epsilon)\left(R\frac{\pd R_\epsilon}{\pd R}\right)+\int_0^{R_\epsilon}\left(R\frac{\pd {\cal A}}{\pd R}-{\cal A}\right).
\end{equation}

As an illustration let us briefly discuss the application of this procedure for pure $AdS_4$. For this geometry equation (\ref{extremal}) yields
\be
\tilde r(\rho)= \frac{l_4^2}{ \sqrt{R^2-\rho^2}}.
\ee
This corresponds to an extremal surface reaching into the bulk until $r=l_4^2R^{-1}.$
Using equation (\ref{reg2}) we find
\be
R_\epsilon= \sqrt{ R^2-\frac{\epsilon^2}{2}}.
\ee
For this simple case we can compute equation (\ref{ree}) explicitly. First we find 
\be
\mathcal{A} (R,\rho)= \frac{R \rho} {(R^2-\rho^2)^{3/2}},
\ee
and then 
\be
\int_0^{R_\epsilon} d\rho\, {\cal A}(R,\rho)= \frac{R}{\sqrt{R^2-R_\epsilon^2}}-1,
\ee
which diverges as we take $\epsilon\rightarrow 0$. However, the additional terms in (\ref{ree}) contribute as follows:
\beq
{\cal A}(R,R_\epsilon)\frac{\pd R_\epsilon}{\pd R}&=&\frac{R^2}{(R^2-R_\epsilon^2)^{3/2}} \cr
\int_0^{R_\epsilon} d\rho\, \frac{\pd{\cal A}}{\pd R}&=&-\frac{R_\epsilon^2}{(R^2-R_\epsilon^2)^{3/2}}.
\eeq
So we find that for $AdS_4$ 
\be\label{AdS REE}
{\cal S}(R) =1.
\ee
This result is consistent with the interpretation of the REE as a c-function probing the holographic renormalization group flow. 
%%%%%%%%%%%%%%%%%%%%
%%%%%%%%%%%%%%%%%%%%
%%%%%%%%%%%%%%%%%%%%
\section{BPS black objects in $AdS_4$}\label{BPS}

We wish to apply the techniques presented in the previous section to an interesting class of solutions of the form (\ref{metric ansatz}), namely  $\frac{1}{4}$-BPS black objects in ${\cal N}=2$, FI gauged supergravity. These correspond to zero temperature solutions supported by scalars and abelian gauge fields. 
They are parametrized by two vectors of $2 n_v+2$  real quantities
\begin{equation}\label{parameters}
 \Gamma= \left(\begin{array}{c}p^I \\
                                        q_I \end{array}\right)
\hspace{7mm}\text{and}\hspace{7mm}
G= \left(\begin{array}{c}g^I \\
                                        g_I \end{array}\right),
\end{equation}
where $p^I$ and $q_I$ are the magnetic and electric charges of the gauge fields while $g^I$ and $g_I$ are the parameters of the Fayet-Iliopoulos potential. In (\ref{parameters}) the index $I=1,\dots, n_v+1$, where $n_v$ is the number of vector multiplets considered.  These parameters are not independent. In fact, they are subject to the symplectic constraint
\begin{equation}
\langle G , \Gamma \rangle =k,
\end{equation}
where just as in (\ref{metric ansatz}), $k$ labels the horizon topology, e.g., $k=-1,0,1$ indicates spherical, flat and hyperbolic respectively.

In the following discussion we focus only on the properties of the spacetime metric and leave the behavior of the scalars and gauge fields aside. In terms of the quantities (\ref{parameters}), the warp factors are given by \cite{Katmadas, Halmagyi:2014qza}
\begin{equation}\label{warps}
e^{\psi}=a\,b=\left(I_4(G)^{1/4}\,r+\langle G,B\rangle\right) r\hspace{8mm} b=\frac{1}{2}\, I_4(H)^{1/4},
\end{equation}
where $I_4(V)$ is defined in equation (\ref{I4 1}) in the Appendix, and $H$ is a symplectic vector of linear functions
\begin{equation}
   H=A\,r+B.
\end{equation}
Here, $A$ and $B$ are constant symplectic vectors. The former can be obtained directly from the FI parameters
\begin{equation}
   A=\frac{1}{2} I_4(G)^{-3/4} dI_4(G),
\end{equation}
while the latter is given by a combination of the charges and FI parameters, dictated by the algebraic equation
\begin{equation} \label{find b} 
\frac{1}{4}dI_4(B,B,G)=\Gamma,
\end{equation}
where $dI_4$ is defined in equation (\ref{dI}) in the Appendix. 
Moreover, physical consistency requires to choose solutions of (\ref{find b}) that fulfill the constraints
\begin{equation}\label{physical conditions}
 \langle G,B\rangle>0 \hspace{10mm} I_4(B)>0.
\end{equation}
Therefore the construction of a BPS solution is reduced to a purely algebraic problem.

The warp factors (\ref{warps}) correspond to a metric that interpolates between $AdS_4$ at infinity (UV) and an $AdS_2\times \Sigma_k^2$ near horizon (IR) geometry as $r \rightarrow 0$. The AdS radii corresponding to these UV and IR geometries are given by 
\begin{equation}\label{radii}
    l_{4}=I_4(G)^{-1/4} \hspace{10mm}    l_{2}=\frac{1}{2}I_4(B)^{1/4}\langle G,B\rangle^{-1}.
\end{equation}
Moreover, the entropy (density for k=0) is proportional to 
\begin{equation}\label{sigma 0}
b^2\xrightarrow{\hspace{2mm} IR\hspace{2mm}} \sigma_0^2= \frac{1}{4}\sqrt{I_4(B)}.
\end{equation}
At this point we want to remind the reader that these solutions are also accompanied by flowing scalars.  Due to the attractor mechanism \cite {Dall'Agata:2010gj, Barisch:2011ui} the scalars flow from constant to constant. These scalars can be thought of as coupling constants and giving rise to the notion of an attractive RG flow \cite{WITS}. 
%%%%%%%%%%%%%%%%%%%%
%%%%%%%%%%%%%%%%%%%%
%%%%%%%%%%%%%%%%%%%%
\section{REE for BPS black branes} \label{branes}
%%%%
In this section we compute the renormalized entanglement entropy as discussed in section (\ref{REEa}) for solutions of the kind presented in section \ref{BPS}. In the following computation we restrict ourselves to black objects with flat horizons ($k=0$), i.e., black branes. Furthermore, we will consider solutions of the STU-model. This model captures the essential features of extremal black holes in ${\mathcal N} \geq 2, d=4$ theories \cite {Bellucci:2008sv}. In the STU-model, the structure constants for the prepotential (see Appendix \ref{ApA}) are given by $c_{ijk}= |\epsilon_{ijk}|$. By plugging the warp factors (\ref{warps}) into equation (\ref{extremal}), we obtain an explicit, albeit complicated, differential equation for $r$ in terms of $\rho$. Now we need to set the parameters that will support the solution i.e. charges/FI parameters. To each such charge/FI configuration we can associate three symplectic invariant combinations, which correspond to the AdS length scales and the entropy density. As we will see in the following our results depend only on these three quantities. In order for the solution to be regular these quantities must not vanish and we choose the charges/FI parameters accordingly. 
%%%%%%

Hereafter we will consider solutions supported by non-vanishing charges/FI parameters
\begin{equation}\label{charge flux}
(q_0, p^i ; g^0, g_i)\hspace{9mm} i=1,\dots, n_v+1
\end{equation}
or
\begin{equation}
(p^0, q_i ; g_0, g^i)\hspace{9mm} i=1,\dots, n_v+1.
\end{equation}
In the following discussion we will display the results for the first configuration (\ref{charge flux}). For the second configuration we have verified explicitly that we get completely analogous results. It is straightforward to extend the following discussion to other configurations as well. Moreover, the reader must keep in mind that the results that follow are invariant under symplectic transformations of the kind discussed in \cite{deWit:2001pz, Bellucci:2008cb}.

\begin{figure}[h!]
  \centering
\includegraphics[trim=1cm 1.5cm 0cm 0cm,scale=0.42]{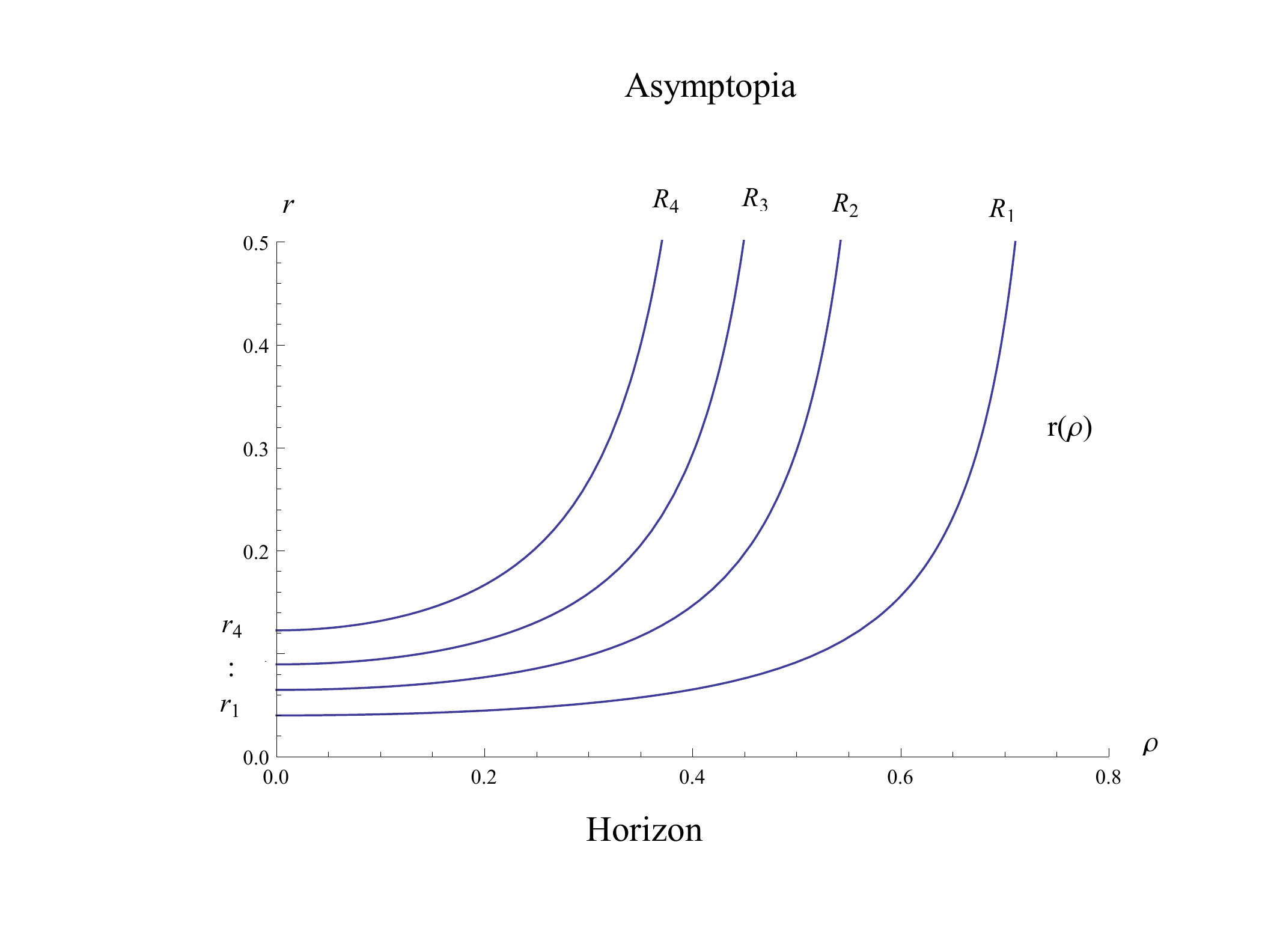}
  \caption{Some extremal surfaces for a $p^1=2$, $p^2=p^3=1$ and $-g^0=g_1=g_2=g_3=1$ black brane}\label{minimal numeric}
\end{figure}

Given the intricacy of the resulting ODE describing the extremal surfaces' profile, we proceed to solve it numerically. Moreover, in order to realize the program outlined in section \ref{REEa} we are compelled to produce a large sample of such minimal surfaces $r_i(\rho)$ with $i$ being an index for the sample (see Fig. \ref{minimal numeric}). Now by introducing a cut-off $\epsilon$ it is possible to compute numerically (\ref{area functional reg}), creating thus a list of regularized areas $S_i(\epsilon)$ corresponding to each of the extremal surfaces $r_i(\rho)$. Hence we are left with a list of points $(R_i, S_i(\epsilon))$, which can be interpolated to find 
$S(R,\epsilon)$. Finally, from this function we construct the desired renormalized entanglement entropy ${\cal S} (R)$ for a given set of charges/FI parameters. We must point out that by construction ${\cal S} (R)$ is a cut-off  independent quantity. This behavior is exhibited by our numerical computations as we tune $\epsilon$ to ever smaller values. 

\begin{figure}[h!]
  \centering
\includegraphics[trim=3.5cm 3.5cm 1cm 3cm,scale=0.5]{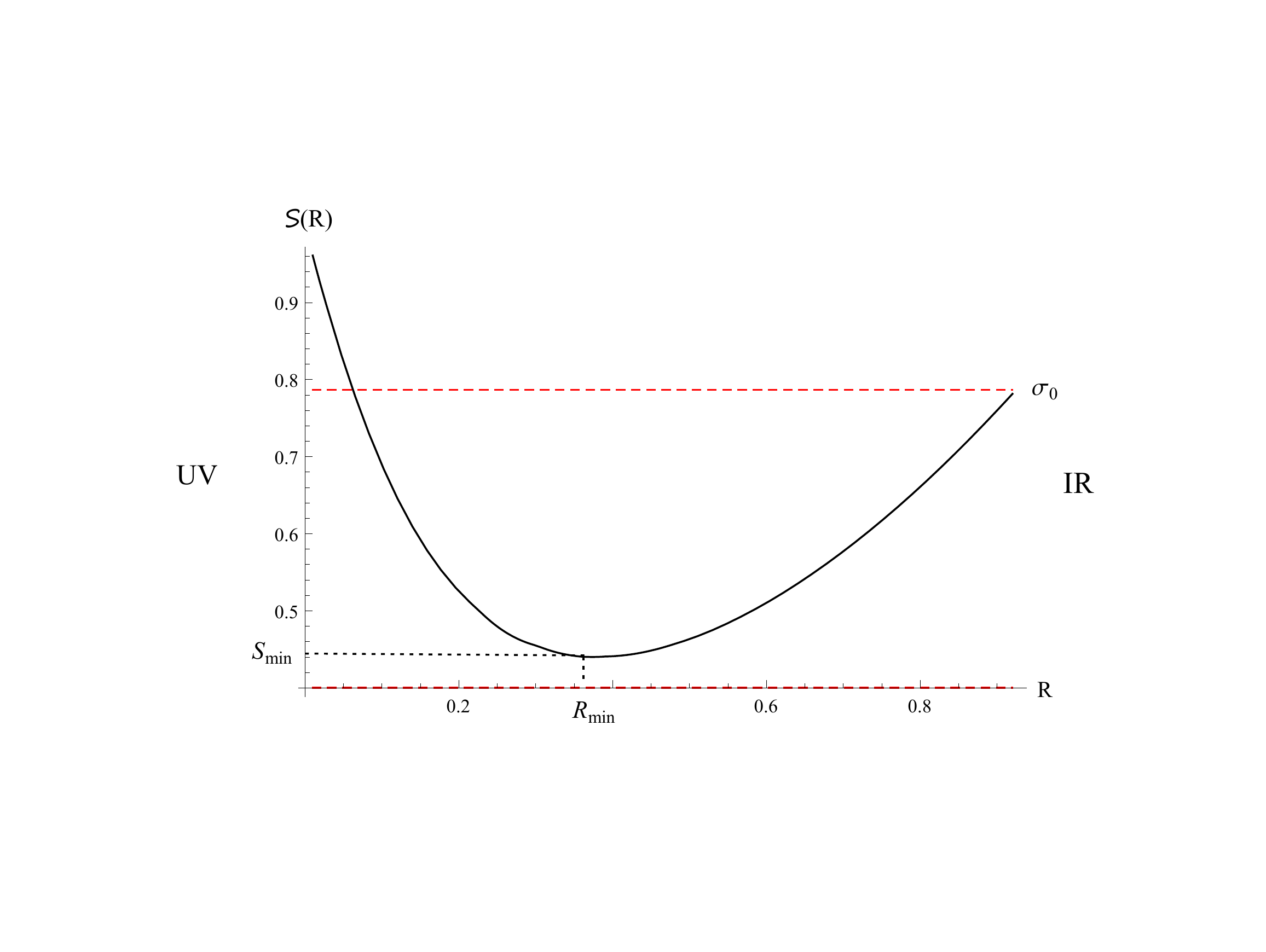}
  \caption{Renormalized entanglement entropy for a $p^1=2$, $p^2=1$, $p^3=1/2$ and $-g^0=g_1=g_2=g_3=1$ black brane}\label{REE}
\end{figure}

The resulting REE is depicted in Fig. \ref{REE} for a particular example. However the observed behavior is generic regardless of the values  chosen for the parameters (\ref{charge flux}). First of all, as  ${\cal S} (R)\rightarrow 1$ as $R\rightarrow 0 $ in agreement with (\ref{AdS REE}), the REE then decreases monotonically until it reaches a minimum ${\cal S}_*$ when the entangling disk has a radius $R_*$. After reaching that critical value,  ${\cal S} (R)$ starts to increase and approaches the value $\sigma_0$ as we get closer to the horizon. 

We wish to explore how the values of ${\cal S}_{*}$ and $R_{*}$  depend upon the charges/FI parameters $\Gamma$ and $G$.
First of all, it is clear that these parameters must enter only through symplectic invariant combinations. 
The warp factors (\ref{warps}) can be specified in terms of the invariant quantities $\{ I_4(G), \langle G,B \rangle, I_4 (A), I_4(B)  \}$. From equation (\ref{radii}, \ref{sigma 0}) we can identify them as $\{ l_2, l_4, \sigma_0, I_4(A)\}$. But a closer inspection shows that $I_4(A)$ can be expressed in terms of $l_4$. Hence, there are three independent invariants upon which ${\cal S}_{*}$ depends, namely the AdS radii and the entropy density. 
%%%%%%%%
%%%%%%%%%
 In order to find a pattern, we start by identifying a subclass of parameters for which one of the symplectic invariants is held fixed. 
One such family is given by
\be\label{t3 charges}
g_1=g_2=g_3 \hspace{10mm}   p^1=p^2=p^3.
\ee
The crucial point here is that for black branes with these kind of parameters the near-horizon $AdS_2$ radius $l_2$ is independent of the value of $p^1$. Let's see how this comes about. For  charge/FI combinations of the form (\ref{t3 charges}) the solution of equation (\ref{find b}), consistent with the conditions (\ref{physical conditions}), reads
\be
B^1=B^2=B^3=\lambda_1\,\text{sgn} (g^0)\sqrt{\frac{p^1 }{g_1}}, \ B_0=\lambda_2\,|g^0|^{-1} \sqrt{p^1\,g_1}.
\ee
Here $\lambda_1$ and $\lambda_2$ are known positive constants, and the components of $B$ that are omitted vanish. The upshot is that $B$ is proportional to $\sqrt{p^1}$. Therefore, since $l_2$ in (\ref{radii}) is invariant under rescalings of $B$, it is clear that the $p^1$ dependence drops out. 

Now we fix the overall scale $l_4$, and study the behavior of ${\cal S}_{*}$. 
In this context, changing $p_1$ gives rise to a one parameter family of solutions with constant $l_2$ and varying $\sigma_0$. Interestingly, we find that ${\cal S}_{*}$ is constant along this family, which implies that ${\cal S}_{*}$ is a function of $l_2$ only. 
%%%%%%
Notice that in the regime $R=0$ to $R=R_*$ the REE decreases monotonically from the pure $AdS_4$ value to a constant which can be determined 
solely from the $AdS_2$ radius. This is reminiscent of the c-function discussed in \cite{WITS}. It would be interesting to investigate this connection further. 
%%%%%%

\begin{figure}[h!]
 \centering
\includegraphics[trim=1cm 1.5cm 1cm 1cm,scale=0.50]{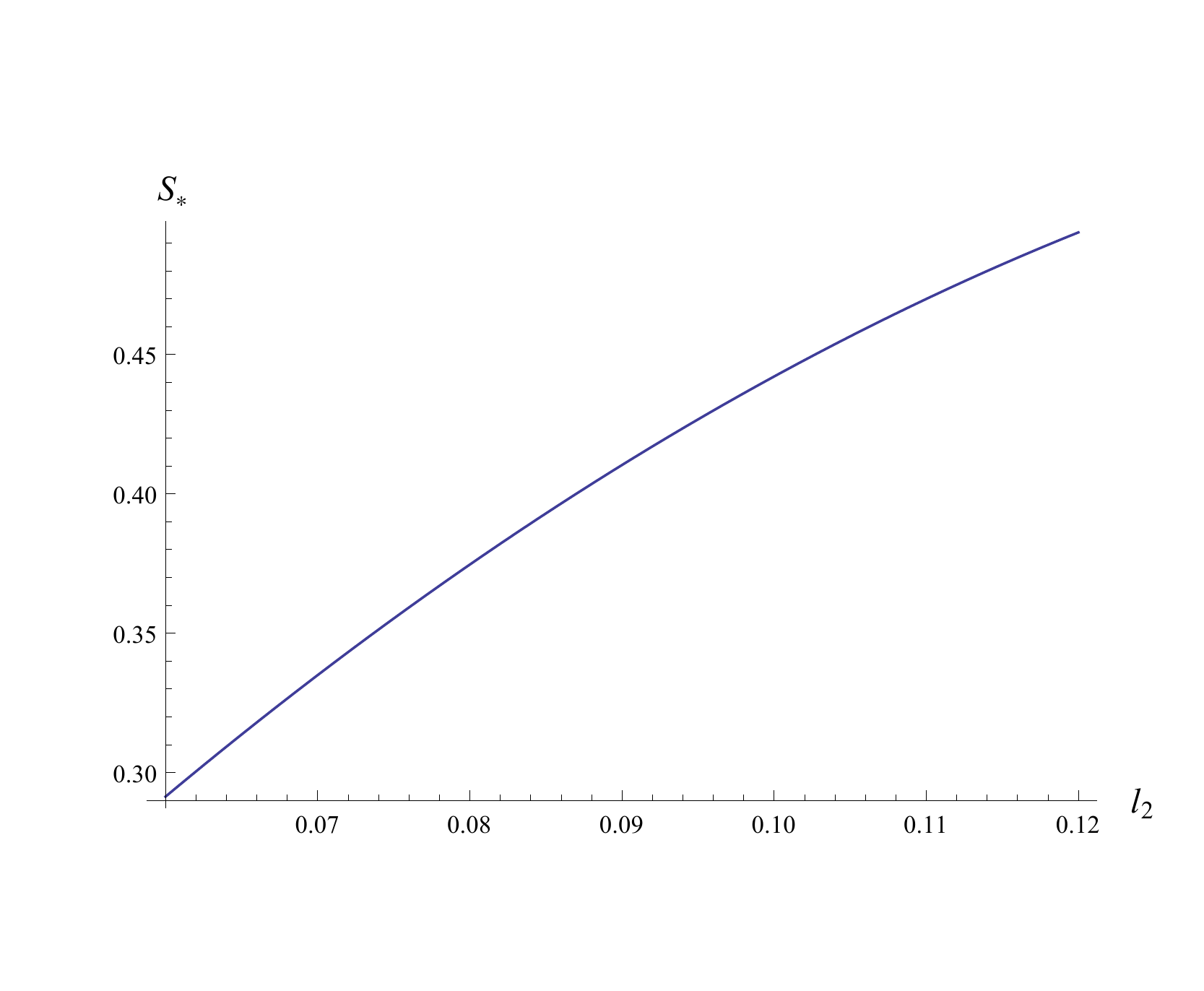}
\caption{Variation of ${\cal S}_{*}$ with $l_2$ }\label{Smin}
\end{figure}
Then we explore how ${\cal S}_{*}$ varies with $l_2$. The variation is displayed in Fig (\ref{Smin}). 
%%%%%%%%
This plot clearly shows that ${\cal S}_{*}$ increases with $l_2$. On the other hand, $R_*$ depends on both $l_2$ and $\sigma_0$. Moreover, it increases with $l_2$ and decreases with $\sigma_0$. 
%%%%%%%%%%%%%%%%%%%%%%
%%%%%%%%%%%%%%%%%%%%%%
%%%%%%%%%%%%%%%%%%%%%%
\section{Discussion} 
In this paper we have computed the renormalized entanglement entropy, ${\cal S}$ for 4d, ${\cal N}=2$ BPS black brane solutions. These solutions interpolate between $AdS_4$ in the UV and a space with an $AdS_2$ factor in the IR. Specifically, we have investigated the behavior of ${\cal S}$ as a function of the size of the entangling region. We have found that in this context ${\cal S}$ first decreases monotonically with $R$, reaches an extremum and then increases again. This is the key finding of our investigation.

We have already mentioned in the main text that all the parameters of the brane solutions enter by three independent symplectic invariant combinations that can be identified as the three independent scales of the system, namely $l_4, l_2$ and $ \sigma_0$. In our inspection we found that starting from the UV the ${\cal S}$ monotonically decreases until it reaches a minimum that is determined completely by the radius of $AdS_2$. Then it starts to increase again and approaches the black brane entropy density. Furthermore, as pointed out in the previous section, when we increase the entropy of the brane this transition occurs closer to the UV. The fact that the ${\cal S}$ decreases monotonically for that region in R resembles the behavior of a c-function for vacuum states. Then the chemical potential starts to dominate once we go deeper in the IR. 

Entanglement entropy measures the entropy due to tracing out part of the total system. If the total system is in a mixed state this quantity receives contributions both from entanglement and from the mixedness of the original system. Since black branes correspond to mixed states we expect our computation to be influenced by both of these factors. In light of that, we are inclined to interpret our result in the following way--the REE is driven predominantly by entanglement close to the UV before reaching $R_*$, where contributions due to the mixedness of the branes take over. 

Since there are three scales in the system it is amiable to find that the REE (with R) probes these scales and passes through those critical points.
It would be interesting to see if we can use the behavior of REE as a function of R in order to establish an order parameter for the phase transition between the vacuum to vacuum flow and vacuum to charged state flow. This is because the existence of the extremum for the REE tells us that the system is transiting from its vacuum behavior at that point. 
%%%%%%%%%%%%%%%%%
%%%%%%%%%%%%%%%%%
%%%%%%%%%%%%%%%%%
\section*{Acknowledgements}
We are grateful to Michael Haack for commenting on a draft of this paper. We would  like to thank Kevin Goldstein, Kiril Hristov, Vishnu Jejjala, Stefanos Katmadas, Mark Mezei, Suresh Nampuri and Aninda Sinha for helpful conversations. The research of SSH and AVO is supported by the University Research Council and the National Research Foundation. SSH would like to thank Perimeter Institute for their hospitality where part of the work was done. AVO is grateful to the theory group at Universit\`a di Milano-Bicocca for kindly hosting him during the concluding stages of this work. AB is grateful to the string theory group at Harish-Chandra Research Institute, Allahabad, India, where part of this work was completed.\\
%%%%%%%%%%%%%%%%
%%%%%%%%%%%%%%%%
%%%%%%%%%%%%%%%%
%%%%%%%%%%%%%%%%
%%%%%%%%%%%%%%%%
%%%%%%%%%%%%%%%%
\appendix
\section{Duality transformations for ${\cal N}=2$ gauged supergravity}\label{ApA}	
Supergravity solutions can be easily written in terms of \textit{symplectic vectors}. These are vectors with $2 n_v+2$ components of 
which the first $n_v+1$  components are labeled with an upper index and the remaining ones
with a lower index, e.g.,
\begin{equation}
V=\left(\begin{array}{c}V^I\\V_I
    \end{array}\right).
\end{equation} 
$I=0,\dots,n_v$.
These vectors are acted upon by symplectic transformations
\be
\left(\begin{array}{c} \tilde V^I\\ \tilde V_I
    \end{array}\right)
=
\left(\begin{array}{cc} \ U & Z\\  W & V
    \end{array}\right)
\left(\begin{array}{c}V^I\\  V_I
    \end{array}\right),
\ee
where
\bea
 U^T V-W^TZ&=&V U^T-WZ^T=1 \nonumber\\
U^TW=W^TU&,& Z^TV=V^TZ .
\eea
We refer to these reparametrizations as duality transformations \cite{deWit:2001pz}.

In ${\cal N}=2$ models, physical quantities must be invariant under duality transformations. Duality invariant quantities can be succinctly expressed in terms of the symplectic quartic invariant $I_4$ \cite{Ferrara:1997uz}
\begin{eqnarray}\label{I4 1} \nonumber
I_4\left(V\right) &=&\frac{1}{4!}t^{MNPQ}V_M\,V_N\,V_P\,V_Q\\ \nonumber
&=&-\left(V^IV_I\right)^2+\frac{2}{3}V_0\,c_{ijk}V^iV^jV^k-\frac{2}{3}V^0\,c^{ijk}V_iV_jV_ k \\
&+&c_{ijk}\,c^{lmn}V^iV^jV_lV_m.
\end{eqnarray}
In this expression and the ones to follow, the lower-case indices run from $1$ to $n_v$ only. The constant coefficients $c^{ijk}$ encode the underlying special geometry prepotential 
\begin{equation}
F=-\frac{1}{6}c_{ijk}\frac{X^i X^jX^k}{X^0}.
\end{equation}
 For future convenience, we also define 
\begin{equation} \label{dI}
dI_4(V)= \Omega_{MN}\frac{\partial I_4(V)}{\partial V_N},
\end{equation}
with 
\begin{equation}
 \Omega_{MN}= \left(\begin{array}{c}\hspace{1mm}0\hspace{4mm} \mathbb{I} \\
                                      - \mathbb{I}  \hspace{4mm} 0\end{array}\right),
\end{equation}
the canonical symplectic matrix.
Moreover, given four symplectic vectors we define %\footnote{Notice the absence of the overall symmetrization factor wrt (\ref{I4 1}). {\textit{This seems to be the usual convention for some reason}}}
\begin{equation}
   I_4\left(V^{(1)},V^{(2)},V^{(3)},V^{(4)}\right)= t^{MNPQ}\,V_M^{(1)}\,V_N^{(2)}\,V_P^{(3)}\,V_Q^{(4)}.
\end{equation}
Notice the absence of the overall symmetrization factor with respect to (\ref{I4 1}).
In practice, we can obtain the $t$-tensor by hitting (\ref{I4 1}) with four derivatives. 
The black hole solution is determined by a set of electric/magnetic charges and fluxes.
%%%%%%%%%%%%%%%%%%%%
%%%%%%%%%%%%%%%%%%%%
%%%%%%%%%%%%%%%%%%%%

\end{document}